\documentclass{emulateapj}
\usepackage{apjfonts}
\usepackage{epsf}
\begin{document}
\slugcomment{}
\shortauthors{J. M. Miller et al.}
\shorttitle{NGC 1275}


\title{Chandra Imaging of the Outer Accretion Flow onto the Black Hole at the Center of the Perseus Cluster}

\author{J.~M.~Miller\altaffilmark{1},
M.~W.~Bautz\altaffilmark{2}, 
B.~R.~McNamara\altaffilmark{3,4} 
}

\altaffiltext{1}{Department of Astronomy, University of Michigan, 1085
  South University Avenue, Ann Arbor, MI 48109-1107, USA,
  jonmm@umich.edu}

\altaffiltext{2}{Kavli Institute for Astrophysics \& Space Research,
Massachusetts Institute of Technology, 77 Massachusetts Avenue,
Cambridge, MA 02139, USA}

\altaffiltext{3}{Department of Physics and Astronomy, University of
Waterloo, 200 University Avenue West, Waterloo, ON N2L 3G1, Canada}

\altaffiltext{4}{Perimeter Institute for Theoretical Physics,
Waterloo, ON N2L 2Y5, Canada}

\begin{abstract}
Nowhere is black hole feedback seen in sharper relief than in the
Perseus cluster of galaxies.  Owing to a combination of astrophysical
and instrumental challenges, however, it can be difficult to study the
black hole accretion that powers feedback into clusters of galaxies.
Recent observations with {\it Hitomi} have resolved the narrow Fe
K$\alpha$ line associated with accretion onto the black hole in NGC
1275 (3C 84), the active galaxy at the center of Perseus.  The width
of that line indicates the fluorescing material is located 6--45~pc
from the black hole.  Here, we report on a specialized {\it Chandra}
imaging observation of NGC 1275 that offers a complementary angle.
Using a sub-array, sub-pixel event repositioning, and an X-ray ``lucky
imaging'' technique, {\it Chandra} imaging suggests an upper limit of
about 0.3 arc seconds on the size of the Fe K$\alpha$ emission region,
corresponding to $\sim$98 pc.  Both spectroscopy and direct imaging
now point to an emission region consistent with an extended molecular
torus or disk, potentially available to fuel the black hole.  A low
X-ray continuum flux was likely measured from NGC 1275;
contemporaneously, radio flaring and record-high GeV fluxes were
recorded.  This may be an example of the correlation between X-ray
flux dips and jet activity that is observed in other classes of
accreting black holes across the mass scale.
\end{abstract}

\section{Introduction}
NGC 1275 is the galaxy at the heart of the Perseus Cluster
($z=0.017284$, Hitomi Collaboration 2017); accordingly, it harbors a
massive very black hole ($M_{\rm BH} = 3.4\times 10^{8}~M_{\odot}$,
Wilman et al.\ 2005; $M_{\rm BH} = 8\times 10^{8}~M_{\odot}$,
Scharwachter et al.\ 2013).  Accretion onto this black hole fuels
strong feedback into the intracluster medium (ICM), shaping the bulk
of the baryonic matter within the gravitational potential (e.g.,
Fabian et al.\ 2003, 2006).  Indeed, the Perseus Cluster is arguably
the most dramatic and accessible example of strong feedback between
massive black holes and large-scale structure.  Currently, at least,
the black hole in NGC 1275 radiates at a low fraction of its Eddington
limit ($L/L_{\rm Edd} \sim 3\times 10^{-4}$, Sikora et al.\ 2007),
below the quasar and Seyfert regimes.

Jet-driven shocks and bubbles are apparent in many clusters (e.g.,
Virgo/M87: Churazov et al.\ 2001, Forman et al.\ 2007; Hydra A:
McNamara et al.\ 2000).  Feedback is also implied by the lack of very
cold gas and star formation expected based on observed cooling
luminosities (e.g., Peterson et al.\ 2003, Sanders et al.\ 2010),
indicating some manner of energy injection.  Mechanisms such as
turbulence and conduction may be important (e.g., Peterson \& Fabian
2006); however, recent results from {\it Hitomi} find that the ICM is
not highly turbulent within Perseus (Hitomi Collaboration, 2016).

To fully understand strong feedback, it is also important to study the
accretion modes of the massive black holes that drive the process.
Especially in the midst of complex continuum emission, the clearest
X-ray signature of AGN activity is the presence of a strong, neutral
Fe~K$\alpha$ line at 6.40~keV.  This line is produced through the
irradiation of cool, optically thick material by hard X-ray emission
from the accetion process (e.g., Lightman \& White 1988, George \&
Fabian 1991).  Narrow line emission is likely tied to the illumination
of the broad line region (BLR) and/or the distant molecular torus
inferred in unification models and detected in some IR, sub-mm, and
radio observations (e.g., Burtscher et al.\ 2013, Scharwachter et
al.\ 2013, Fuller et al.\ 2016).

The Fe K$\alpha$ line in NGC 1275 has been studied in several prior
observations, notably by {\it XMM-Newton} (Churazov et al.\ 2003).
The resolution of CCD detectors is too coarse to measure the velocity
width of such a line.  However, recent observations of Perseus with
{\it Hitomi} have detected the Fe~K$\alpha$ line from NGC 1275.  A
FWHM of 500--1600~km/s is measured, consistent with a molecular torus
or extended molecular disk (Hitomi Collaboration, 2017).  In this
work, we present complementary {\it Chandra} imaging and spectroscopy
of NGC 1275.\\

\section{Observations and Reduction}
We observed NGC 1275 with {\it Chandra} for a total of 100~ks,
comprised of four individual exposures.  The AGN is brighter in the
last three observations, rendering them less suited to the subject of
this investigation.  In this work, we focus on ObsID 19568, which
observed NGC 1275 starting on 1 November 2016, for a total exposure of
32.4~ks.  The source was placed at the default aimpoint of the ACIS-S3
chip.  A 1/8 sub-array was implemented to reduce the nominal frame
time from 3.2~s to 0.4~s, in order to limit photon pile-up.  After
standard processing, the net exposure time of the delivered "evt2"
file is 28.5~ks.

The data were manipulated using the tools and packages available
through the standard CIAO suite, version 4.9, and the associated
calibration files.  Images were examined using the "ds9" package.
Spectral analysis was performed using XSPEC version 12.9 (Arnaud
1996).  The $\chi^{2}$ statistic was minimized in spectral fits, using
the standard weighting scheme.  The errors reported in this work
reflect the values of a parameter at its $1\sigma$ confidence
limits.\\

\section{Analysis and Results}

\subsection{Initial Analysis}
We initially examined "evt2" images of NGC 1275 covering the standard
ACIS band (0.3-8 keV).  Using the CIAO tool "dmcopy", we binned the
full "evt2" image of NGC 1275 to have a pixel resolution 10 times
sharper than the native resolution (0.0495 arc seconds versus 0.495
arc seconds).  This is possible because {\it Chandra} dithers, and the
intensity of a source can be tracked as it passes over pixel
boundaries.  The ability of this technique to separate closely-spaced
sources and to sharpen the image of those sources has been
demonstrated in numerous papers (e.g., Li et al.\ 2004; Wang et
al.\ 2009, 2011).

We initially inspected the full "evt2" image at this improved spatial
resolution; the image does not reveal strong photon pile-up.  However,
the effects of pile-up are still evident in the spectrum.  A number of
different source and background extractions (made using the CIAO
"specextract" tool) measure implausibly hard power-law photon indices
in the spectrum ($\Gamma \simeq 0.5$), with stronger positive
residuals at higher energy.  This is characteristic of spectra
adversely affected by pile-up.  In such cases, the source image can be
artificially broadened.

{\it Chandra}/ACIS events are assigned a "grade", based on the pattern
of the charge within a 3x3-pixel event box.  "Good" events are those
consistent with a photon depositing charge into just a few pixels.
Plausible source spectra produce recognized patterns of event grades;
conversely, some distributions of event grades signal problems.  One
way of flagging photon pile-up is through ``grade migration'': at high
count rates, multiple photons are more likely to strike within a
single event box, within a single frame time.  These events may still
register as ``good'' events, but the charge is more likely to be
distributed in several pixels (Davis 2001; also see Miller et
al.\ 2010).

In order to further mitigate photon pile-up effects, then, we used
"dmcopy" to create a ``grade=0'' event list.  These events are
single-pixel photon strikes.  The resultant image and
spectrum are as free as possible from imaging and spectral distortions
due to photon pile-up.  Accepting only single-pixel ACIS events in
this circumstance is similar to ``lucky imaging'' in optical
photometry (e.g., Law et al.\ 2006).  The resolution achieved by
combining all of these procedures is illustrated in Figure 1.

\subsection{Spectroscopy}
As noted above, an initial spectral extraction made using the full set
of ``good'' event grades suffers from photon pile-up distortions;
the Fe K$\alpha$ line cannot be recovered in the associated spectra.
We therefore extracted source and background events from the
``grade=0'' file.  The CIAO tool ``wavdetect'' was run to determine
the source extent.  At native pixel resolution, and using the full
energy band, ``wavdetect'' finds a slightly elliptical annulus, with
semi-axes of 1.96 and 1.84 arc seconds.  Background
events were extracted using a circular annulus centered on the source,
and spanning 3-5 arc seconds.  The CIAO tool ``specextract'' was then
run to create binned source and background spectra, and associated
response files.

We fit the ``grade=0'' spectrum with a simple power-law function in
the 2.3-9.0~keV band.  This reaches slightly above the band typical of
ACIS spectroscopy, but many HETG spectra are fit up to 10~keV, and the
extra range is helpful to isolating the AGN continuum.  Above 9~keV,
the effects of photon pile-up are more pronounced.  Below 2.3~keV, the
source is strongly dominated by diffuse emission from the cluster.
Thus, the chosen band enables a focus on emission from the AGN itself.
The best-fit power-law index trends toward $\Gamma = 1.1$, but it is
clear that this is driven by a small degree of residual photon pile-up
above approximately 7~keV.  The low-energy part of the specrum is
better fit by a power-law with a typical photon index.  We therefore
set a lower bound of $\Gamma = 1.4$, consistent with the limits of
Comptonization (e.g., Haardt \& Maraschi 1993), and slightly flatter
than reported in other recent work (e.g., $\Gamma = 1.55$, Churazov et
al.\ 2003).  The data prefer a fit at the limit, giving $\Gamma =
1.40^{+0.02}$.  This continuum model returns a fit statistic of
$\chi^{2}/\nu = 69.3/52 = 1.33$ (see Figure 2).

The addition of an unresolved (FWHM$=$0 keV) Gaussian function
measures a line energy of E$=6.16^{+0.04}_{-0.08}$~keV, and a flux of
F$=3.6\pm 1.2 \times 10^{-6}~{\rm photons}~{\rm cm}^{-2}~{\rm
  s}^{-1}$.  This improves the fit statistic to $\chi^{2}/\nu =
61.0/50.0 = 1.22$, making the improvement significant at the 95\%
level of confidence as measured with an F-test.  However, especially
for single emission lines, it is more appropriate to measure
significance using the line normalization and its errors; this gives a
significance of $3\sigma$.  The line equivalent width is W$=300\pm
100$~eV.

This significance of the line is modest in these data, but the
detection is likely to be robust.  The observed line energy of
E$=6.16^{+0.04}_{-0.08}$~keV corresponds to
E$=6.27^{+0.04}_{-0.08}$~keV in the host frame.  After accounting for
plausible gain uncertainties (see Section 4.2), the measured energy is
consistent with the rest-frame energy of $E_{\rm lab} = 6.40$~keV.

For completeness, we also examined the sensitivity of the line
equivalent width to the continuum.  Again, the true continuum cannot
be as hard as $\Gamma = 1.1$, but fits with this continuum give
W$=240\pm 90$~eV.  If the continuum is fixed at $\Gamma = 1.55$, as
per the prior {\em XMM-Newton} observation (Churazov et al.\ 2003),
the line equivalent width increases to W$=360\pm 110$~eV.  These
values are formally consisent with our preferred treatment of the
continuum with $\Gamma \geq 1.4$ set as a floor.

Finally, we note the presence of an apparent line at approximately
4.45~keV.  The feature is nominally significant at the $\sim2\sigma$
level of confidence.  However, this feature cannot be associated with
an He-like or H-like charge state of any abundant element in the host
frame, even allowing for gain uncertainties (see below).  After
accounting for the trials in a blind search through the grouped
spectral bins, the true significance of the feature is just over
1$\sigma$.

\subsection{Spectroscopy Checks}
In order to evaluate the effect of ``grade=0'' filtering, we examined
a long {\it Chandra} observation of the supernova remnant N103b.  This
source is fairly compact (its radius is approximately 16 arc seconds),
and regular (see, e.g., van der Heyden et al.\ 2002), and it is bright
enough to establish flux offsets.  Most importnatly, the strong atomic
lines in its spectrum make evaluations of the energy scale possible.
{\it Chandra} obsID 1045 started on 17 January 2001; the net
integration time was 73.9~ks.  The observation was made using the
HETG, which had the fortuitous effect of limiting photon pile-up in
the zeroth-order data that we analyzed.  In addition to the standard
``evt2'' event list including all ``good'' event grades, we extracted
a ``grade=0'' event list.  For both, source events were extracted from
a 15 arc second circle, and background events were extracted from a
source-free region closeby.  The ``specextract'' tool was run to
create binned source and background spectra and responses.

The spectra were jointly fit in XSPEC, using an ``apec'' model with
variable abundances.  An overall constant was allowed to float between
the spectra, to measure any flux offsets.  All fits to the spectra
require that the ``grade=0'' spectrum be multiplied by a value of
0.30-0.33.  This indicates that ``grade=0'' filtering may artificially
reduce the flux of a source by a factor of $\sim3$, relative to a
reduction that includes all ``good'' ACIS grades (Li et al.\ 2004
reach the same conclusion for the ACIS-I detector).  We note that this
is effectively a lower limit in the case of NGC 1275, since the source
is piled-up.  The ``apec'' model includes a velocity shift parameter;
allowing the redshift of the ``grade=0'' spectrum to vary gives a
value of $v\simeq0.015c$ (this is strongly required; $\Delta\chi^{2} =
85$ for one interesting parameter), or about 0.1~keV at 6.4~keV.
Thus, the line detected in our spectrum of NGC 1275 is consistent with
the Fe~K$\alpha$ line at 6.4~keV.

The particular advantage of N103b is that it is compact and is not
highly susceptible to gain drifts across read-out nodes on the ACIS
chips.  However, as a second check on the performance of the detector
when ``grade=0'' filtering is applied, we extracted the spectrum of
the diffuse cluster gas in larger regions within the 1/8 sub-array.
As detailed best by {\it Hitomi}, the diffuse cluster gas contains
strong emission lines, especially the He-like Fe XXV complex.  This
can potentially serve as a second gauge of the gain calibration.

At very high resolution, the Fe XXV complex is a sequence of four
lines; at lower resolution, it is a complex of three; at CCD
resolution, and with modest sensitivity, it is a single feature that
can be fit by a simple Gaussian.  The strongest line in the complex is
the resonance line; in an ``apec'' plasma, the line centroid is
6.705~keV.  It is 3--4 times stronger than any other line in the
complex, and fits with a single Gaussian should be strongly weighted
toward its centroid energy.  Fits to spectrum of the diffuse cluster
gas after ``grade=0'' filtering give a line centroid of
$6.51\pm0.02$~keV.  This corresponds to $6.62\pm0.02$~keV in the
source frame, or about 0.07--0.11~keV below the Fe XXV resonance line.
This broadly confirms the results of our first check with N103b, and
supports our association of the line at E$=6.16^{+0.04}_{-0.08}$~keV
in the observed frame with the 6.40~keV Fe K$\alpha$ line in the host
frame.

\subsection{Imaging}
Figure 1 shows the full 0.3--8.0~keV image of NGC 1275, following
``grade=0'' filtering, sub-pixel processing and Gaussian smoothing.
We ran the standard CIAO tool "wavdetect" on this image using the
default settings.  This tool detects a slightly elliptical
source, with semi-axes of 0.49 and 0.46 arc seconds.
The broader, low-level emission surrounding the source is not
axially symmetric, and the flux decrements to the south and northeast
appear to correspond to dust lanes in Hubble images; a full analysis
of the combined {\it Chandra} and {\it Hubble} images will be treated
in a separate paper.

At the distance of NGC 1275, 1 arc second corresponds to a physical
size of approximately 357~pc.  Averaging the semi-axes of the source
region would give a radial extent of 170~pc.  Functionally, this
extent is an upper limit.  It is possible that the AGN is still hidden
within the more diffuse emission of the ICM.  We therefore filtered
the event list to only select photons in the 6.0--6.4~keV band,
covering the Fe~K$\alpha$ line.  In this image, ``wavdetect'' finds a
source with semi-axes of 0.31 and 0.24 arc seconds.  The bright core
of the image is quite symmetric.  Averaging these axes gives an
effective upper limit on the extent of the source of 98~pc.  Figure 3
shows the image of NGC 1275 in the Fe K band, and the detected source
region.

The Fe~K band image of NGC 1275 does not contain only line photons; it
also contains continuum photons.  It is possible that the Fe K
emission region is larger (or, slightly resolved), while the true
continuum emission is only point-like.  To test this possibility, we
created an image in the 5.2--5.6~keV band.  In this continuum reference
band, ``wavdetect'' returns a source with semi-axes of 0.35 and 0.24
arc seconds.  Thus, there is no evidence that the line region is any
larger than the continuum region.

\subsection{Imaging Checks}
In order to check the validity of our imaging, we ran
simulations using the {\it Chandra}/MARX suite, version 5.3.2.
Numerous runs were set-up, including point sources and Gaussian
sources with different physical extent (e.g. $\sigma = 0.2, 0.3, 0.5$
arc seconds, and so on).  The parameters of our observation --
including the source position, boresight, count rate, and exposure
time, etc. -- were replicated.  Our simulations also used the
``dithering'' functionality within MARX, in order to later enable
sub-pixel imaging.  The ``marx2fits'' tool was used to create fake
event lists from the MARX simulations.

In simulations that assumed a point source, ``wavdetect'' recovered a
source with semi-axes of 1.0--1.1 arc seconds in event lists including
all ``good'' event grades, and binned at the native pixel resolution.
The source semi-axes are reduced to 0.4 arc seconds, after sub-pixel
event repositioning (0.0495 arc second pixels, as per our analysis of
obsID 19568).  When ``wavdetect'' was then run on only ``grade=0''
sub-pixel images, the source semi-axes were reduced to sizes as small
as 0.2 arc seconds.  In no point source simulations were sizes smaller
than 0.2 arc seconds recovered.  We note that image sizes smaller than
0.2 arc seconds should not be recovered in actual data, based on the
level of blurring incurred because of finite limits on the accuracy
with which the observatory aspect can be tracked\footnote[1]{See
  Section 4.4 of the Chandra Proposer's Observatory Guide, available
  at http://asc.harvard.edu/proposer/POG.}.  In the Fe~K band, our
detected source sizes are only slightly larger than this minimum, and
a small degree of photon pile-up remains, so we regard the source as
unresolved.


\section{Discussion}
%

Jets from massive black holes in the central galaxies of clusters
drive feedback into the intracluster medium, shaping the bulk of the
baryonic matter in the largest gravitationally bound structures in the
universe.  The complex environment in clusters and their central
galaxies, coupled with modest X-ray fluxes from black holes radiating
well below their Eddington limit, complicates observational efforts to
understand the nature of the disks and accretion flows that launch
such powerful feedback.  Narrow, low-ioniation Fe~K$\alpha$ lines are
a nearly ubiquitous feature of Seyfert AGN, and likely arise through
X-ray irradiation of the broad line region and/or the molecular torus
(e.g. Nandra et al.\ 2007, Shu et al.\ 2010).  These
lines are sufficiently distinct to utilize as accretion flow
diagnostics even in the complex environments and spectra seen in
cluster centers.

We have therefore analyzed a special {\it Chandra} observation of the
massive black hole in NGC 1275 (3C 84), which drives strong feedback
into the Perseus cluster.  A combination of strategies were employed
to obtain the sharpest possible spatial resolution.  Our analysis of
the 6.0--6.4~keV image of NGC~1275 suggests that the Fe~K$\alpha$
emission line region is likely smaller than about 98~pc in size,
consistent with an extended molecular torus or disk in NGC 1275
(Wilman et al.\ 2005, Sharwachter et al.\ 2013), representing the
outermost accretion flow.  A recent analysis of the X-ray calorimeter
spectrum of NGC 1275 obtained with {\it Hitomi} measures an
Fe~K$\alpha$ line FWHM of 500--1600~km/s, or 6--45~pc assuming
Keplerian rotational broadening for plausible black hole masses
(Hitomi Collaboration, 2017).  The sharpest possible X-ray imaging and
spectroscopy point to a similar size scale and toroidal origin for the
Fe~K$\alpha$ line in NGC 1275.  This implies that a Seyfert-like
geometry may still hold in NGC 1275, though the black hole has a
radiative Eddington fraction of just $3\times 10^{-4}$ (Sikora et
al.\ 2007) and launches powerful jets.

Using the relationships developed by Reynolds et al.\ (2000), the Fe~K
line equivalent width that we have measured implies a gas mass of
$M\simeq 6\times 10^{8}~M_{\odot}$, for solar metallicity and a unity
covering factor.  Although our limit of 98~pc is a few times larger
than plausible estimates of the Bondi radius of the central black hole
in NGC 1275, it is now very clear that a large reservoir of cold gas
sits closeby, and may be available for accretion.  The reservoir is
nominally sufficient to power the black hole at its Eddington limit
for an AGN lifetime of $\simeq10^{7}$ years (assuming an efficiency of
10\% for disk accretion), or in its current mode for about $10^{4}$
times longer.  Estimates of the jet power based on ICM bubbles
($L_{kin} \simeq 10^{44}~{\rm erg}~{\rm s}^{-1}$; Dunn \& Fabian 2004)
may require the mass fueling rate to be higher by a factor of several
over that implied by the radiative luminosity, assuming the efficiency
for conversion of the accretion rate into radiation and kinetic power
is the same.  It is worth noting that advanced imaging with {\it
  Chandra} has resolved the Bondi capture radius around particularly
massive and/or nearby black holes in simpler environments (e.g.,
Baganoff et al.\ 2003; Wong et al.\ 2011, 2014).

Through an analysis of the supernova remnant N103b and the diffuse
cluster gas in Perseus, we have attempted to understand the systematic
effects of ``grade=0'' filtering.  A small gain shift appears to be
required, and absolute fluxes may be low by a factor of three.  The
Fe~K$\alpha$ line flux that we measured is formally consistent with
the {\it Hitomi} measurement (Hitomi Collaboration, 2017), but an
upward revision would make it more consistent with prior {\it
  XMM-Newton} measurements (Churazov et al.\ 2003).  The uncertainties
on all line flux measurements in the literature are fairly large and
there is broad agreement.  The line equivalent width we have measured,
${\rm W} = 300\pm 100$~eV, is an order of magnitude higher than
recorded by {\it Hitomi} (though only a factor of $\sim2$ lower than
measured with {\it XMM-Newton}).  The change in equivalent width is
likely driven by a low continuum flux.  Whereas {\it Hitomi} measured
a 2--10~keV flux of $2.05^{+0.44}_{-0.54}\times 10^{-11}~ {\rm erg}~
{\rm cm}^{-2}~ {\rm s}^{-1}$, our fits give a flux of just
$1.0_{-0.1}^{+0.05}\times 10^{-12} ~ {\rm erg}~ {\rm cm}^{-2}~ {\rm
  s}^{-1}$.  This could be a manifestation of the X-ray Baldwin effect
(see, e.g., Iwasawa \& Taniguchi 1993).  Examination of N103b
suggests that at least a third of the continuum deficit can likely be
attributed to ``grade=0'' filtering.

Contemporaneously with our {\it Chandra} obsevation, NGC 1275
exhibited remarkable behaviors in radio and MeV bands, consistent with
enhanced jet activity.  The MAGIC telescopes recorded ``giant
flaring'' activity above 100~GeV from NGC 1275 on 29 October 2016
(Mirzoyan et al.\ 2016).  On 30 October 2016, VERITAS recorded the
highest $170$~GeV flux ever measured from NGC 1275, approximately four
times higher than prior flares (Mukherjee et al.\ 2017). Thoughout
November, 2016, the RATAN-600 radio telescope recorded NGC 1275 in a
high flux state (45~Jy at 11~GHz; Trushkin, Nizhelskij, \& Tsybulev
2016), elevated by a factor of two above the 2~GHz flux density
measured a year prior.

In stellar-mass black holes, quasars, and even in low-luminosity AGN
(LLAGN), flux dips in the X-ray band are associated with enjanced jet
activity (e.g., Fender \& Belloni 2004; Chatterjee et al.\ 2009, 2011;
King et al.\ 2016).  The simplest explanation of such phenomena is
that the inner accretion flow is ejected in such episodes.  It is
possible, then, that the low continuum flux that we have observed is a
flux dip associated with jet activity in NGC 1275.  It is also
possible that the radio and GeV behaviors relate to downstream
activity in the jet, largely disconnected from the central engine.
Future X-ray observations triggered based on the radio and/or GeV flux
levels of cluster AGN may help to reveal accretion-ejection coupling
in these key sources.

JMM acknowledges helpful conversations with Ping Zhao, Patrick Slane,
Randall Smith, and the {\it Chandra} helpdesk.  We thank Joel Kastner
for discussions regarding the mechanics and validity sub-pixel event
repositioning and grade filtering.  We are grateful to the anonymous
referee, for comments improved this manuscript.

\clearpage

\begin{figure}
  \hspace{0.5in}
  \includegraphics[scale=0.85]{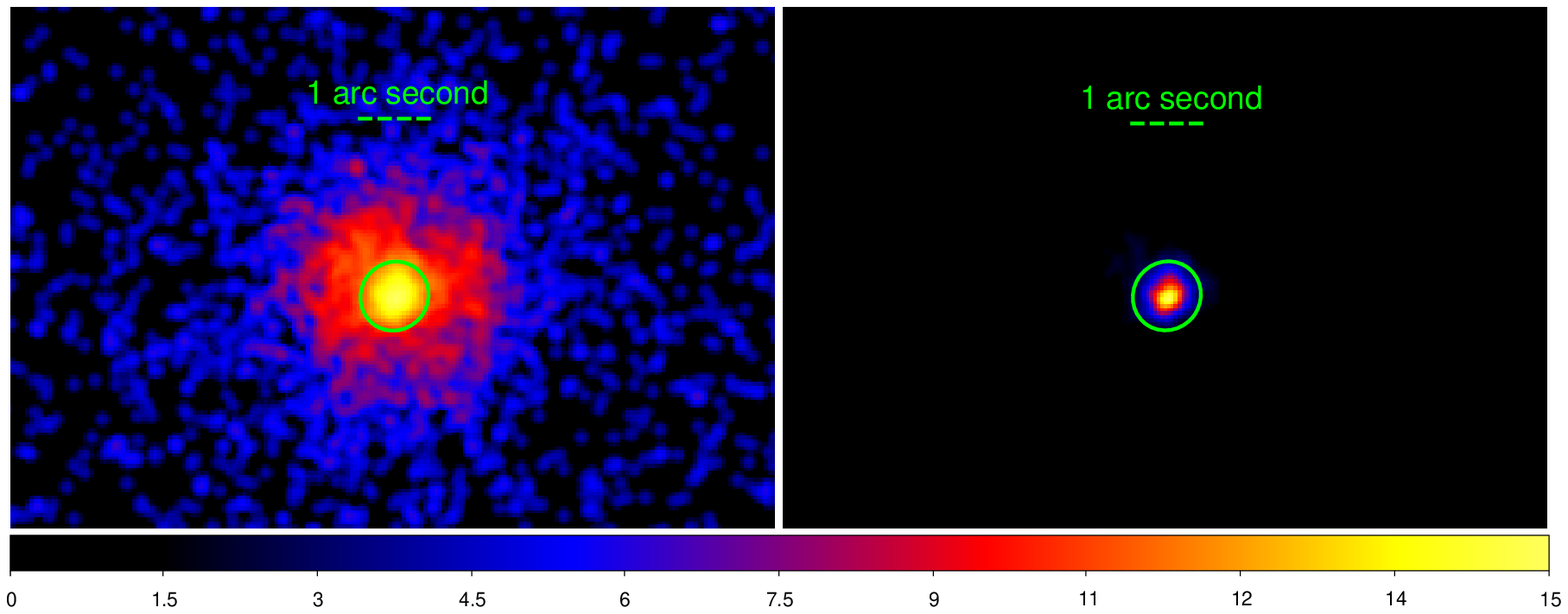}
  \figcaption[t]{\footnotesize Images of NGC 1275 in the 0.3--8.0~keV
    band from our 1/8 sub-array exposure.  The data were filtered to
    only include ``grade=0'' events; the image was binned to a
    resolution of 0.1 native pixels (0.0495 arcseconds), and smoothed
    using a 3-subpixel Gaussian.  The left-hand image has a
    logarithmic color stretch.  The right-hand image is shown using a
    linear color stretch.  In both images, the CIAO ``wavdetect''
    source region is shown; it is an elliptical Gaussian with
    semi-axes of 0.49 and 0.46 arc-seconds, and the source is easily
    encompassed within the region.  Averaging these semi-axes
    corresponds to a radius of 169~pc at the distance of NGC 1275.}
\end{figure}
\medskip

\begin{figure}
  \hspace{1.0in}
  \includegraphics[scale=0.5,angle=-90]{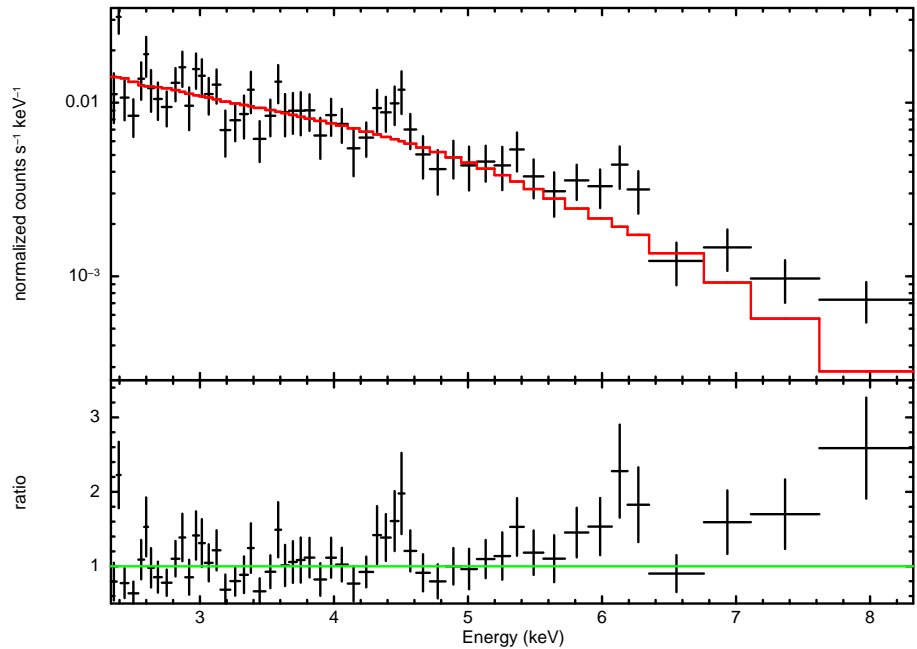}
  \figcaption[t]{\footnotesize The {\it Chandra} ``grade=0'' ACIS-S
    spectrum of NGC 1275 in the observed frame.  The spectrum is only
    fit above 2.3~keV in order to focus on power-law emission from the
    AGN; below this energy, the spectrum is dominated by 
    diffuse emission from the ICM.  The model shown above is a
    power-law with a photon index of $\Gamma = 1.4$, consistent with
    prior studies of NGC 1275.  An upward trend in the data/model
    ratio above approximately 7~kev indicates a degree of residual
    photon pile-up.  The emission line at 6.2~keV is the Fe K$\alpha$
    line at 6.4~keV in the AGN frame (allowing for a modest gain
    correction).  The line is significant at the $3\sigma$ level of
    confidence; its flux is broadly consistent with other recent
    detections but its equivalent width is higher, likely owing to a
    drop in the continuum flux level.}
\end{figure}
\medskip

\begin{figure}
  \hspace{0.5in}
  \includegraphics[scale=0.85]{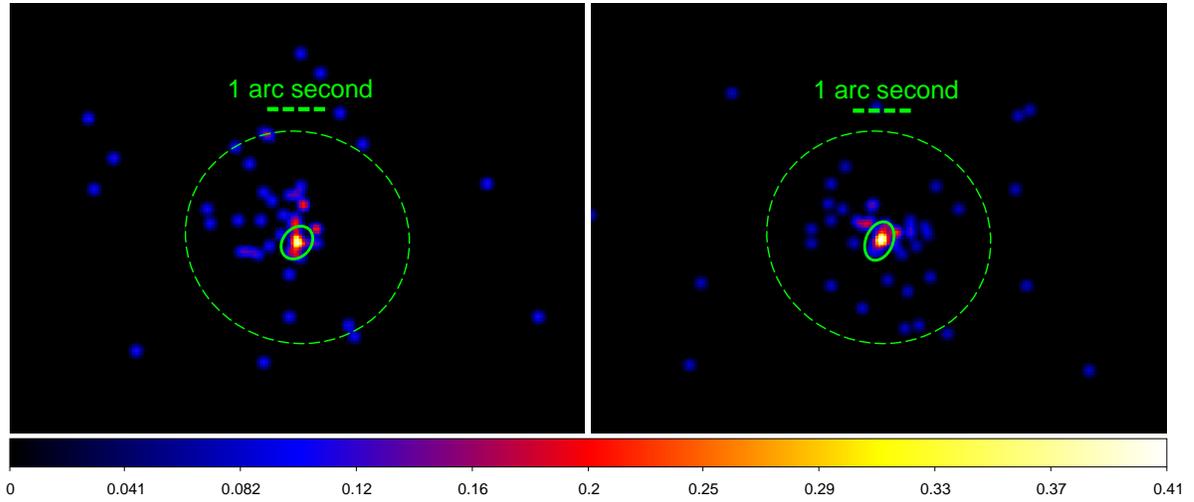}
  \figcaption[t]{\footnotesize Images of NGC 1275 in specific energy
    bands, following ``grade=0'' filtering, sub-pixel processing, and
    smoothing by 3 (sub-) pixels for visual clarity.  The small
    ellipse indicates the source extent determined in each narrow
    band, determined using ``wavdetect''.  The larger dashed ellipses
    indicate the source size returned by ``wavdetect'' when operating
    on the full energy band, all ``good'' event grades, and the native
    pixel size.  LEFT: The 6.0--6.4~keV (Fe K$\alpha$ band) image of
    NGC 1275; the source region has semi-axes of 0.31 and 0.24 arc
    seconds.  RIGHT: The 5.2--5.6~keV image of NGC 1275, a
    continuum--only reference band for the Fe K$\alpha$ image.  The
    source region has semi-axes of 0.35 and 0.24 arc seconds.  The
    line region is not more extended than a nearby continuum band,
    signaling that the line region is likely not resolved.  A
    reasonable upper-limit for the size of the Fe K$\alpha$ line
    emission region is 98~pc, based on an average of the
    semi-axes. Please see the text for details.}
\end{figure}
\medskip

\end{document}